\documentclass[a4paper,12pt]{article}  
\usepackage{amssymb} 

\usepackage{comment} 

\def \ccomma{\raise 2pt\hbox{,}} 
\def \D {\hbox{d}}

\def \diag{\mathop{\rm diag}\nolimits}
\def \cc {C}
\def \uplus {\bar u}
\def \vplus {\bar v}

\def \PVI    {{\rm P_{\rm VI}}} 
\def\qPVI   {{\rm q-P_{\rm VI}}}
\def\auu{a}
         
\begin{document}

\title{Holomorphic, rational Lax pairs of a $q$-discrete Painlev\'e VI equation}

\author{Robert Conte${}^{1,2,3}$     
{}\\
\\ 1. Universit\'e Paris-Saclay, ENS Paris-Saclay, CNRS
\\    Centre Borelli, F-91190 Gif-sur-Yvette, France
\\
\\ 2. Department of mathematics, The University of Hong Kong,
\\ Pokfulam, Hong Kong
\\
\\ 3. Centre de Recherches math\'ematiques, Universit\'e de Montr\'eal, 
\\ C.P. 6128, Succ. Centre-ville, Montr\'eal H3C 3J7, Qu\'ebec, Canada
\\						
\\    E-mail Robert.Conte@cea.fr,         ORCID https://orcid.org/0000-0002-1840-5095
{}\\
}

\maketitle

\hfill 

{\vglue -10.0 truemm}
{\vskip -10.0 truemm}

\begin{abstract}
We build several matrix Lax pairs of $\qPVI$
valid even when the two eigenvalues of the residue of the monodromy matrix at infinity are equal.
Their elements are rational functions of the dependent variables.
\end{abstract}

\noindent \textit{Keywords and phrases} 

Discrete Painlev\'e equations, Bonnet surfaces, discrete Lax pairs


 

												
\noindent \textit{MSC 2020} 
33E17, 
37J70, 
53A05  

\baselineskip=12truept %


\tableofcontents

\vfill \eject

\section{Introduction: $\qPVI$}

The $q$-discrete system
\cite[Eqs (19)--(20)]{JS1996} 
(notation $u=u(x), \uplus=u(q x), v=v(x), \vplus=v(q x)$)
\begin{eqnarray}
& &
\left\lbrace
\begin{array}{ll}
\displaystyle{
u \uplus=a_3 a_4 \frac{(\vplus-a_1 a_2 x/t_1)(\vplus-a_1 a_2 x/t_2)}{(\vplus-1/(\kappa_1 q))(\vplus-1/(\kappa_2))}\ccomma
}\\ \displaystyle{
v \vplus =\frac{1}{\kappa_1 \kappa_2 q} \frac{(u-a_1 x)(u-a_2 x)}{(u-a_3)(u-a_4)}\ccomma
}\\ \displaystyle{
t_1 t_2=\kappa_1 \kappa_2 a_1 a_2 a_3 a_4,
}
\end{array}
\right.
\label{eq-qPVI-JS}
\end{eqnarray}
in which the eight parameters $t_1, t_2, \kappa_1, \kappa_2, a_1, a_2, a_3, a_4$
are complex constants,
admits a doubly discrete second order Lax pair \cite{JS1996} with a spectral parameter $z$
($A$ and $B$ second order matrices),
\begin{eqnarray}
& &
\left\lbrace
\begin{array}{ll}
\displaystyle{
\Psi(x q,z)=B(x,z) \Psi(x,z), \Psi(x, z q)=A(x,z) \Psi(x,z),
}\\ \displaystyle{
X \equiv B(x,z q) A(x,z) - A(x q,z) B(x,z)=0,
}
\end{array}
\right.
\label{eq-q-Lax}
\end{eqnarray}
and a continuum limit to the generic $\PVI$ equation,
\begin{eqnarray}
& & {\hskip -20.0 truemm}
\frac{\D^2 u}{\D x^2}=
 \frac{1}{2} \left[\frac{1}{u} + \frac{1}{u-1} + \frac{1}{u-x} \right] \left(\frac{\D u}{\D x}\right)^2
- \left[\frac{1}{x} + \frac{1}{x-1} + \frac{1}{u-x} \right] \frac{\D u}{\D x}
\nonumber \\ & & {\hskip -20.0 truemm} \phantom{123456}
+ \frac{u (u-1) (u-x)}{2 x^2 (x-1)^2}
  \left[\theta_\infty^2 - \theta_0^2 \frac{x}{u^2} + \theta_1^2 \frac{x-1}{(u-1)^2}
        + (1-\theta_x^2) \frac{x (x-1)}{(u-x)^2} \right],
\nonumber \\ & & {\hskip -20.0 truemm}
(2\alpha, -2\beta, 2\gamma, 1-2\delta)=(\theta_\infty^2,\theta_0^2,\theta_1^2,\theta_x^2),
\label{eqPVI}
\end{eqnarray} 
two properties which make it deserve the name $\qPVI$ system.

Already written explicitly in \cite[Eq (9)]{RGH1991}, 
with a fourth order matrix Lax pair presented in \cite{GNPRS1994},
it was only recognized as a $\qPVI$ equation
after the discovery of the discrete Lax pair of (\ref{eq-qPVI-JS}).

The goal of the present work is to remove two unpleasant features 
of the Lax pair of Ref \cite{JS1996}, namely
\begin{enumerate}
	\item 
	the restriction to unequal values of $\kappa_1$ and $\kappa_2$,
	\item 
	a nonrational dependence on $(u,\uplus,v,\vplus)$ of the off-diagonal elements of the Lax pair
	(only their discrete logarithmic derivative is rational).
\end{enumerate}

Indeed, these two features, which are already present in the original matrix Lax pair 
of the continuous $\PVI$ \cite{JimboMiwaII},
can be removed by converting the moving frame of Bonnet surfaces
to a second order, traceless, matrix Lax pair
\cite{Conte-Lax-PVI-CRAS} 
\cite{Conte-Bonnet-JMP}.
In the present work, we also remove them for $\qPVI$.

The paper is organized as follows.

In section \ref{sectionSchlesinger}
we recall the prescriptions of Schlesinger for a second order matrix Lax pair of (continuous) $\PVI$,
and point out the unfortunate consequences for both $\PVI$ and $\qPVI$ of a suggested simplification.

In section \ref{section-PVI-Lax-various},
we recall two different implementations of the prescriptions of Schlesinger 
yielding for $\PVI$ a matrix Lax pair at the same time holomorphic in the 
monodromy exponents and rational in $u(x), u'(x),x$.

In section \ref{section-Holomorphic},
making a nondiagonal matrix assumption instead of a diagonal one,
we obtain fifteen rational, holomorphic Lax pairs for $\qPVI$,
and we present explicitly two of them.

\vfill\eject

\section{$\PVI$. The prescriptions of Schlesinger for a matrix Lax pair}
\label{sectionSchlesinger}

In the matrix case, the simplest nontrivial monodromy is defined
by a second order matrix whose Fuchsian singularities
are four nonapparent ones and zero apparent \cite{SchlesingerP6}.
This is an advantage over the scalar case,
in which the simplest nontrivial monodromy arises from 
a second order linear ordinary differential equation
containing four nonapparent and one apparent Fuchsian singularities \cite[p 219]{Poincare1884}. 

Neither Poincar\'e nor Schlesinger performed the associated computations.
As is well known, this was achieved for the first time
in the scalar case by Richard Fuchs \cite{FuchsP6} (the brother-in-law of Schlesinger),
and in the matrix case by Jimbo and Miwa \cite{JimboMiwaII}.

However, this matrix Lax pair displays the same two unpleasant features 
(meromorphic dependence on one parameter, nonrational elements) than the one of $\qPVI$,
and this is for the same reason, so let us first point out this reason,
then explain how to restore a holomorphic dependence on the monodromy exponents
and a rational dependence for the matrix elements.

The Lax pair of Schlesinger has the structure ($x$ denotes the independent variable, $t$ the spectral parameter),
\begin{eqnarray}
& & {\hskip -15.0 truemm}
\D \Psi=L \Psi \D x + M \Psi \D t,\ 
\nonumber \\ & & {\hskip -15.0 truemm}
M=\frac{M_0}{t}+\frac{M_1}{t-1}+\frac{M_x}{t-x},\
L=                             -\frac{M_x}{t-x}+L_\infty,\
M_\infty=-M_0-M_1-M_x,
\label{eq-Matrix-Lax-pair}
\end{eqnarray}
in which the four residues $M_j$ and the regular term $L_\infty$ only depend on $x$
and can be chosen traceless.

In particular, $L_\infty$ is not necessarily zero,
and $M_\infty$ is not necessarily constant.
However, Schlesinger points out \cite[p 105]{SchlesingerP6} a simplification,
namely the existence,
\textit{on the condition} that the residue at infinity $M_\infty$ be invertible,
of a transition matrix able to, at the same time, make $M_\infty$ diagonal and constant,
and make $L_\infty$ zero.

Implementing this suggested simplification,
Jimbo and Miwa assumed
\begin{eqnarray}
& & 
(\PVI):\ M_\infty=\hbox{constant}=\diag(a_\infty/2,-a_\infty/2), 
L_\infty=0, 
\end{eqnarray} 
and the condition $a_\infty\not=0$
(otherwise the monodromy matrix would admit less than four Fuchsian singularities)
led them to the restriction $\theta_\infty^2 \not=1$,
where $\theta_\infty$ is one of the four monodromy exponents, defined 
in terms of the usual four parameters $\alpha,\beta,\gamma,\delta$ of $\PVI$
by the relations
\begin{eqnarray}
& &
(\theta_\infty^2,\theta_0^2,\theta_1^2,\theta_x^2)= (2 \alpha, -2 \beta, 2 \gamma, 1 - 2 \delta).
\label{eq-theta_j}
\end{eqnarray}
	
Similarly,
the assumption of Jimbo and Sakai, in their notation \cite[Eq (10)]{JS1996},
\begin{eqnarray}
& & 
(\qPVI):\ A_2 = \hbox{constant} = \diag(\kappa_1,\kappa_2),
\end{eqnarray} 
led them to the restriction that the denominator $\kappa_1-\kappa_2$ should not vanish.

\vfill\eject

\section{$\PVI$. The various realizations of a holomorphic matrix Lax pair}
\label{section-PVI-Lax-various}

At least two other realizations of the prescriptions of Schlesinger have been made,
which succeed to remove the meromorphic dependence on one monodromy exponent
and to make the off-diagonal elements rational.
\begin{enumerate}

	\item 
	Assuming a nonconstant never diagonal matrix for the residue at infinity,
	a ternary symmetry for the three other residues,
	and an arbitrary regular term $L_\infty$,
\begin{eqnarray}
             & & {\hskip -9.0 truemm}
M_\infty=\hbox{nondiagonal}=
                                   \pmatrix{ a_\infty/2  & \not=\hbox{constant}  \cr 0  & -a_\infty/2 \cr },
L_\infty=
                \pmatrix{ f_{11}(x)  & f_{12}(x)  \cr f_{21}(x)   & -f_{11}(x)  \cr },
\nonumber \\ & & {\hskip -9.0 truemm}
M_j     =
                \pmatrix{ a_j  & b_j \cr (\theta_j^2/4-a_j^2)/b_j  & -a_j \cr }, j=0,1,x
\hbox{ (invariance 3!)},
\nonumber \\ & & {\hskip -9.0 truemm}
(b_0,b_1,b_x)=\left(-\frac{u}{x},\frac{u-1}{x-1},\frac{u-x}{x(1-x)}\right) \hbox{ so that } M_{12}=\frac{t-u}{t(t-1)(t-x)},
\nonumber \\ & & {\hskip -9.0 truemm}
(a_0,a_1,a_x)=\hbox{lhs of Riccati equations for } u(x),
\nonumber
\end{eqnarray} 
which implies $a_\infty=\theta_\infty -1$,
Frank Loray \cite{Loray2016}	
succeeded to remove the two restrictions,
all matrix elements being quadratic in the four $\theta_j$'s and rational in $u,u',x$.

	\item 
	The other realization also removes the two above restrictions and its advantage
	is to require no assumption at all, thus making it quite natural.
Since Bonnet surfaces \cite{Bonnet1867} have a mean curvature equal to the logarithmic derivative of 
the one-zero tau-function of $\PVI$ \cite[expression $t$ page 341]{ChazyThese}
\cite{BE1998},
their moving frame can be converted to a matrix Lax pair of $\PVI$ 
\cite{Conte-Lax-PVI-CRAS} 
\cite{Conte-Bonnet-JMP},        
with in particular the result that
the residue at infinity is a never diagonal constant matrix
and the regular term $L_\infty$ a scalar multiple of $M_\infty$,
\begin{eqnarray}
             & & {\hskip -9.0 truemm}
 M_\infty=\hbox{constant nondiagonal}=\frac{1}{4}       \pmatrix{ 2 \auu & -4 \cr \auu^2-\theta_\infty^2 & -2 \auu\cr },\
 L_\infty=-\frac{u-x}{x(x-1)}M_\infty.
\nonumber \\ & & {\hskip -9.0 truemm}
\end{eqnarray} 
Moreover, if $\theta_\infty$ is nonzero,
there exists a transition matrix making the Lax pair symmetric with respect to the diagonal
\cite[Eq (42)]{Conte-Bonnet-JMP},
\begin{eqnarray}
 & & M_{12}(\theta_\infty)=M_{21}(-\theta_\infty),
\label{eqPVI-diagonal-symmetry}
\end{eqnarray}         
whose off-diagonal elements are however no more rational.

\end{enumerate}

\vfill\eject
\section{Holomorphic, rational Lax pairs of $\qPVI$}
\label{section-Holomorphic}

If a $q$-discrete version of Bonnet surfaces were known,
the similar conversion of their moving frame would provide
the best $q$-discrete Lax pair of $\qPVI$.
Since this is not (yet) the case,
guided by the holomorphic matrix Lax pairs of $\PVI$ recalled in section \ref{section-PVI-Lax-various},
let us only change the assumption on the residue at infinity from ``constant diagonal'' to ``constant nondiagonal''.

\subsection{Monodromy matrix $A$}

By discretizing the Lax pair of Jimbo and Miwa,
Jimbo and Sakai characterized the $q$-discrete mono\-dromy matrix $A(x,z)$
($z$ spectral parameter, $x$ independent variable)
by the following properties 
\cite{JS1996}
\begin{enumerate}
	\item $A(x,z)=A_2(x) z^2 + A_1(x) z + A_0(x)$;
	\item $A_2(x)=\diag(\kappa_1,\kappa_2)$;
	\item $\forall \lambda:\ \det (A_0(x)-\lambda)=(\lambda-t_1 x)(\lambda-t_2 x)$ ($t_j$ constants);
	\item $\det A(x,z)=\kappa_1 \kappa_2 (z-a_1 x)(z-a_2 x)(z-a_3)(z-a_4)$ ($a_j$ constants),
\end{enumerate}
which imply
\begin{eqnarray} & & t_1 t_2 = \kappa_1 \kappa_2 a_1 a_2 a_3 a_4.
\end{eqnarray}

Instead of assuming $A_2(x)$ constant and diagonal,
which generates a denominator $\kappa_1-\kappa_2$,
let us represent $A_2$ by the never diagonal constant matrix
\begin{eqnarray}
& & {\hskip -15.0 truemm}
A_2(x)=       \pmatrix{ (\kappa_1+\kappa_2)/2+c & -1 \cr c^2-(\kappa_1-\kappa_2)^2/4 & (\kappa_1+\kappa_2)/2-c},
\label{eqA2neverdiagonal}
\end{eqnarray}
in which $c$ is an irrelevant arbitrary constant.
A shift of $c$ will simplify the results,
\begin{eqnarray}
& & \cc=c +\frac{\kappa_1+\kappa_2}{2}\cdot
\end{eqnarray}

One then defines the same three functions of $x$ than Jimbo and Sakai \cite[Eq (14)]{JS1996},
\begin{eqnarray}
& &
\left\lbrace
\begin{array}{ll}
\displaystyle{
u(x)   \hbox{ defined by } A_{12}(x,u)=0,
}\\ \displaystyle{
z_1(x) \hbox{ defined by } A_{11}(x,u)=\kappa_1 z_1,
}\\ \displaystyle{
z_2(x) \hbox{ defined by } A_{22}(x,u)=\kappa_2 z_2,
}
\end{array}
\right.
\end{eqnarray}
definitions which imply the relation
\begin{eqnarray}
& & z_1 z_2 = (u-a_1 x)(u-a_2 x)(u-a_3)(u-a_4).
\end{eqnarray}
Let us introduce the same function $v(x)$ than Ref \cite{JS1996}
\textit{via} the uncoupling,
\begin{eqnarray}
& & z_1 = \frac{(u-a_1 x)(u-a_2 x)}{q \kappa_1 v},
    z_2 =       (u-a_3  )(u-a_4  )  q \kappa_1 v.
\end{eqnarray}

The monodromy matrix is then a Laurent polynomial of 
$A_{1,11}(x)$ and $A_{1,12}(x)$,
\begin{eqnarray}
& & {\hskip -10.0 truemm}
A=\pmatrix{ 
(z-u) A_{1,11} + \cc (z^2-u^2) + \kappa_1 z_1& 
(z-u) [A_{1,12}-z-u] \cr 
A_{21} & 
-(z-u) A_{1,11} - \cc (z^2-u^2) \cr},
\\ & & {\hskip -10.0 truemm}
A_{21}=(z+u) \left[\kappa_1 \kappa_2 (z-u -a_1 x - a_2 x-a_3-a_4) +\cc^2 (z-u)
\right. \nonumber \\ & & {\hskip -15.0 truemm} \phantom{12345} \left.
+ \cc \left\lbrace  (t_1+t_2)x/u+(\kappa_1+\kappa_2)(2u-z)- (\kappa_1 z_1+\kappa_2 z_2)/u \right\rbrace
\right. \nonumber \\ & & {\hskip -15.0 truemm} \phantom{12345} \left.
           +A_{1,11} (2\cc-\kappa_1-\kappa_2)
           +A_{1,12} (\cc-\kappa_1)(\cc-\kappa_2)\right]
        \nonumber \\ & & {\hskip -15.0 truemm} \phantom{12345} 
+ \frac{A_{1,11}}{A_{1,12}} [2(t_1+t_2)x+2(\kappa_1+\kappa_2)u^2-3 \kappa_1 z_1 - \kappa_2 z_2]
+ \frac{A_{1,11}^2}{A_{1,12}} 2 u
        \nonumber \\ & & {\hskip -15.0 truemm} \phantom{12345} 
+ \frac{1}{A_{1,12}} \left[\frac{(\kappa_1 z_1)^2}{u}-(t_1+t_2) x \kappa_1 z_1/u -(\kappa_1+\kappa_2) \kappa_1 z_1 u 
\right. \nonumber \\ & & {\hskip -15.0 truemm} \phantom{12345} \left.
            +\frac{\kappa_1 \kappa_2}{u} (a_1 a_2 a_3 a_4 x^2+(a_3 a_4+(a_1+a_2)(a_3+a_4)x+a_1 a_2 x^2)u^2+u^4)
                      \right].
\end{eqnarray}

constrained by the single algebraic equation
\begin{eqnarray}
& & 
(A_{1,12}- 2 u) A_{1,11}^2 + P_2(A_{1,12}) A_{1,11} + Q_2(A_{1,12})=0,
\label{eqcondA}
\end{eqnarray}
in which $P_2$ and $Q_2$ are second degree polynomials of $A_{1,12}$
with coefficients polynomial in $u,v$ and all the parameters.

Since we want all matrix elements to be rational functions of $u,v$,
let us then require the equation (\ref{eqcondA}) to only admit solutions
$(A_{1,11}(x),A_{1,12}(x))$ rational in $(u,v)$.
This diophantine condition can indeed be solved, as follows.

The discriminant of (\ref{eqcondA}) with respect to $A_{1,11}$, which then must be a square,
has degree two separately in $a_3$ and $a_4$.
therefore its discriminants with respect to $a_3$ and $a_4$ must vanish.
This defines eight values of $A_{1,12}(x)$, 
\begin{eqnarray}
& & 
A_{1,12}=u+a_1 x, u+a_2 x, u+a_3, u+a_4, u, 2 u, u + R_7(u,v), u + R_8(u,v),
\label{eqA112roots}
\end{eqnarray}
with $R_7$, $R_8$ third degree rational functions of $(u,v)$,
and, for each such value, two values (only one for $A_{1,12}=2 u$) of $A_{1,11}(x)$,
presented in the Appendix A.
All fifteen solutions $(A_{1,11}(x),A_{1,12}(x))$ lead to a rational, holomorphic Lax pair,
but, according to the symmetries of the list in Appendix A,
only six of them seem different.

\subsection{Matrix $B$}

In order to take account of (\ref{eqA2neverdiagonal}),
the structure of the matrix 
$B(x,z)$ must be changed from \cite[Eq (12)]{JS1996} to
\begin{eqnarray}
& & 
B=\frac{z(z B_1(x)+B_0(x))}{(z-a_1 q x)(z-a_2 q x)}\cdot
\nonumber
\end{eqnarray}
The commutativity condition (\ref{eq-q-Lax})
then yields the eight elements of $B_1$ and $B_0$
in terms of one of them, for instance $B_{1,12}(x)$, which remains an arbitrary scalar factor,
and it also yields a $\qPVI$ equation characterized by two relations
\begin{eqnarray}
& &
\left\lbrace
\begin{array}{ll}
\displaystyle{
u \uplus=a_3 a_4 \frac{(\vplus-b_1 x)(\vplus-b_2 x)}{(\vplus-b_3)(\vplus-b_4)}\ccomma
}\\ \displaystyle{
v \vplus=b_3 b_4 \frac{(u     -a_1 x)(u     -a_2 x)}{ (u    -a_3)(u     -a_4)}\ccomma
}\\ \displaystyle{
t_1 t_2=\kappa_1 \kappa_2 a_1 a_2 a_3 a_4,
}
\end{array}
\right.
\label{eq-qPVI-neverdiagonal-result}
\end{eqnarray}
in which the $b_j$'s are constants.

Let us explicit two examples.

\vfill\eject
\subsection{First example}


The solution
\begin{eqnarray}
& & 
A_{1,12}=u, A_{1,11}=\frac{\kappa_1 z_1}{u}- t_1 \frac{x}{u}-\cc u,
\label{eq5m}
\end{eqnarray}
is the simplest of the two most symmetric ones.
With the choice $\cc=0$, it yields
\begin{eqnarray}
& & {\hskip -15.0 truemm}
A=   \pmatrix{
\displaystyle    \frac{z \kappa_1 z_1 -(z-u) t_1 x}{u} & 
-z(z-u) \cr 
A_{21} &
\displaystyle    \frac{z \kappa_2 z_2 -(z-u) t_2 x}{u} + (\kappa_1+\kappa_2) z(z-u) \cr},
\nonumber\\ & & {\hskip -15.0 truemm}
A_{21}=\frac{t_1 x \kappa_2 z_2 + t_2 x \kappa_1 z_1}{u^2}
+t_1 x\frac{(\kappa_1+\kappa_2)(z-u)u-2 t_2 x}{u^2}
-(\kappa_1+\kappa_2)\frac{z \kappa_1 z_1}{u}
\nonumber\\ & & {\hskip -15.0 truemm}\phantom{123456}
+\frac{\kappa_1 \kappa_2}{u}\left(z^2+(u-a_1 x-a_2 x-a_3-a_4)z
\right.\nonumber\\ & & {\hskip -15.0 truemm}\phantom{12345678901234}\left.
+(a_1 x+a_2 x)(a_3 a_4 + a_1 a_2 x^2(a_3+a_4))\right).
\end{eqnarray}
and
\begin{eqnarray}
& & 
z(z B_1(x)+B_0(x))= \pmatrix{
\displaystyle (\kappa_1+\kappa_2) z^2- \frac{z^2}{q \vplus} - \frac{t_2 q x z}{u}      + z \frac{a_1 a_2 x^2}{u \vplus}   & z^2 \cr 
 \kappa_1 \kappa_2 z^2 + z \beta_{21} & 
\displaystyle  z\frac{a_1 a_2 x^2 q}{\uplus\vplus}- \frac{z^2}{q \vplus} - \frac{t_1 q x z}{\uplus}   \cr }\ccomma
\nonumber\\ & & 
\beta_{21}=
q u \frac{a_1 a_2 x^2 (3 \kappa_1 \kappa_2 +\kappa_1^2+\kappa_2^2)- (\kappa_1+\kappa_2) (2 t_2 x + \kappa_1 \kappa_2 q a_1 a_2 x^2 \vplus)+t_2^2/(a_1 a_2)  }{t_2 q \vplus - a_1 a_2 x}
\nonumber \\ & &
 \phantom{12}
+\kappa_1 \kappa_2 (a_1+a_2) x q - \frac{\kappa_1 z_1 v}{u \vplus^2}- t_2 \frac{u^2 -a_1 a_2 x^2}{a_1 a_2 x u \vplus}+ (\kappa_1+\kappa_2) \frac{2 u - a_1 x - a_2 x}{\vplus}
\cdot
\label{eqBroot5m}
\end{eqnarray} 

The $\qPVI$ is then
\begin{eqnarray}
& &
\left\lbrace
\begin{array}{ll}
\displaystyle{
u \uplus=a_3 a_4\frac{(\vplus-a_1 a_2 x/t_1))(\vplus-a_1 a_2 x/(t_2 q))}{(\vplus-1/(\kappa_1 q))(\vplus-1/(\kappa_2 q))}\ccomma
}\\ \displaystyle{
v \vplus =\frac{1}{\kappa_1 \kappa_2 q^2} \frac{(u-a_1 x)(u-a_2 x)}{(u-a_3)(u-a_4)}\ccomma
}\\ \displaystyle{
t_1 t_2=\kappa_1 \kappa_2 a_1 a_2 a_3 a_4,
}
\end{array}
\right.
\end{eqnarray}
and it is identical to (\ref{eq-qPVI-JS}) up to a rescaling.

\vfill\eject
\subsection{Second example}


The first four solutions in the list (\ref{eq-list-A112-A111}) are equivalent,
it is therefore sufficient to consider
\begin{eqnarray}
& & {\hskip -15.0 truemm}
A_{1,12}=u+a_1 x, A_{1,11}=\frac{\kappa_1 z_1}{u-a_1 x}-\cc (u+a_1 x).
\label{eq1m}
\end{eqnarray}
With the choice $\cc=0$, one obtains
\begin{eqnarray}
& & 
A=   \pmatrix{
\displaystyle    \frac{\kappa_1 z_1 (z-a_1 x)}{u-a_1 x} & 
-(z-u)(z-a_1 x) \cr 
A_{21} &
\displaystyle    \frac{z \kappa_2 z_2}{u} + (z-u)\left[(\kappa_1+\kappa_2)z - \frac{ a_1 x \kappa_1 z_1}{u(u-a_1 x)} -(t_1+t_2)\frac{x}{u}\right] \cr},
\nonumber\\ & & 
A_{21}=\frac{a_1 x (\kappa_1 z_1)^2}{u(u-a_1 x)^2} 
       -z \kappa_1 \kappa_2 \left[a_2 x+a_3+a_4 - a_2 x a_3 a_4 \frac{z}{u} -u \right] 
\nonumber\\ & & 
\phantom{123456}
       -\frac{\kappa_1 z_1}{u-a_1 x} \left[(\kappa_1+\kappa_2) z -(t_1+t_2) \frac{x}{u} \right]\ccomma
\end{eqnarray}
%
and
\begin{eqnarray}
& & 
z(z B_1(x)+B_0(x))=\pmatrix{
\displaystyle (\kappa_1+\kappa_2) z^2- (t_1+t_2)\frac{q x z}{u} +z \frac{a_1 a_2 x^2 q^2-z u}{qu \vplus}- \frac{z a_1 x q \kappa_1 z_1}{u(u-a_1 x)} & 
z (z-a_1 x q) \cr 
- \kappa_1 \kappa_2 z^2 + z \beta_{21} & 
\displaystyle - \frac{z(z-a_1 x q)}{q \vplus}  \cr }\ccomma
\nonumber\\ & & 
\beta_{21}=(t_1+t_2)\frac{x}{u \vplus} +a_1 x \frac{\kappa_1 z_1}{(u-a_1 x) u \vplus}
                        + a_2 x \left( \kappa_1 \kappa_2 q - \frac{\kappa_1+\kappa_2}{\vplus}+\frac{u-a_1 x q}{q u\vplus^2}\right)\cdot						
\label{eqBroot1m}
\end{eqnarray} 
The $\qPVI$ is then
\begin{eqnarray}
& &
\left\lbrace
\begin{array}{ll}
\displaystyle{
u \uplus=a_3 a_4\frac{(\vplus-a_1 a_2 x/t_1))(\vplus-a_1 a_2 x/t_2)}{(\vplus-1/(\kappa_1 q))(\vplus-1/(\kappa_2 q))}\ccomma
}\\ \displaystyle{
v \vplus =\frac{1}{\kappa_1 \kappa_2 q^2} \frac{(u-a_1 x q)(u-a_2 x)}{(u-a_3)(u-a_4)}\ccomma
}\\ \displaystyle{
t_1 t_2=\kappa_1 \kappa_2 a_1 a_2 a_3 a_4,
}
\end{array}
\right.
\end{eqnarray}
identical to (\ref{eq-qPVI-JS}) up to a rescaling.

\vfill\eject
\section{Conclusion}

It would be interesting to examine
whether one of these fifteen rational, holomorphic matrix Lax pairs 
can, under some change of basis vectors,
acquire the symmetry
\begin{eqnarray}
& & {\hskip -15.0 truemm}
A_{12}(\kappa_1,\kappa_2)=
A_{21}(\kappa_2,\kappa_1),
\end{eqnarray}
i.e.~the discrete counterpart of (\ref{eqPVI-diagonal-symmetry}).
If that would be the case, it would probably be convertible to
the moving frame of a discrete version of Bonnet surfaces.

\section*{Acknowledgements}

The author is pleased to thank the Institute for Mathematical Research of The University of Hong Kong,
and the Institute of Advanced Study of Shenzhen university
for their generous support.

\vfill \eject 

\section{APPENDIX A. Values of $A_{1,11}(x),A_{1,12}(x)$}

For the first six values of $A_{1,12}(x)$ in the list (\ref{eqA112roots}),
the two values           of $A_{1,11}(x)$ are the following
(let us recall that the arbitrary constant $\cc$ can be set to any convenient numerical value),
\begin{eqnarray}
& &
\left\lbrace
\begin{array}{ll}
\displaystyle{
%
A_{1,12}=u+a_1 x, A_{1,11}=\frac{\kappa_1 z_1}{u-a_1 x}-\cc (u+a_1 x),
}\\ \displaystyle{
A_{1,12}=u+a_2 x, A_{1,11}=\frac{\kappa_1 z_1}{u-a_2 x}-\cc (u+a_1 x),
}\\ \displaystyle{
A_{1,12}=u+a_3, A_{1,11}=\frac{\kappa_1 z_1}{u-a_3}-\cc (u+a_3),
}\\ \displaystyle{
A_{1,12}=u+a_4, A_{1,11}=\frac{\kappa_1 z_1}{u-a_4}-\cc (u+a_4),
}\\ \displaystyle{
A_{1,12}=u+a_1 x, A_{1,11}=\frac{\kappa_1 z_1-(t_1+t_2)x}{u} -a_1 x \frac{\kappa_2 z_2}{u(u-a_1 x)}+(\kappa_1+\kappa_2) a_1 x-\cc (u+a_1 x),
}\\ \displaystyle{
A_{1,12}=u+a_2 x, A_{1,11}=\frac{\kappa_1 z_1-(t_1+t_2)x}{u} -a_2 x \frac{\kappa_2 z_2}{u(u-a_2 x)}+(\kappa_1+\kappa_2) a_2 x-\cc (u+a_2 x),
}\\ \displaystyle{
A_{1,12}=u+a_3,   A_{1,11}=\frac{\kappa_1 z_1-(t_1+t_2)x}{u} -a_3   \frac{\kappa_2 z_2}{u(u-a_3  )}+(\kappa_1+\kappa_2) a_3  -\cc (u+a_3  ),
}\\ \displaystyle{
A_{1,12}=u+a_4,   A_{1,11}=\frac{\kappa_1 z_1-(t_1+t_2)x}{u} -a_4   \frac{\kappa_2 z_2}{u(u-a_4  )}+(\kappa_1+\kappa_2) a_4  -\cc (u+a_4  ),
}\\ \displaystyle{
A_{1,12}=u, A_{1,11}=\frac{\kappa_1 z_1}{u}- t_1 \frac{x}{u}-\cc u,
}\\ \displaystyle{
A_{1,12}=u, A_{1,11}=\frac{\kappa_1 z_1}{u}- t_2 \frac{x}{u}-\cc u,
}\\ \displaystyle{
A_{1,12}=2 u, A_{1,11}=\frac{1}{\kappa_1 z_1-\kappa_2 z_2}\left[-\kappa_1 \kappa_2 u (3 u^2-2 (a_1 x+a_2 x+a_3+a_4) u 
\right.}\\ \displaystyle{\left.\phantom{1234567890}
+a_1 x(a_2 x+a_3+a_4)+a_2 x(a_3+a_4)+a_3 a_4)
\right.}\\ \displaystyle{\left.\phantom{1234567890}
 - \frac{t_1 t_2 x^2}{u}
 + (\kappa_1+\kappa_2) \kappa_1 z_1 u
 - (t_1+t_2) x \frac{\kappa_1 z_1}{u}
 + \frac{(\kappa_1 z_1)^2}{u}\right] - 2 \cc u,
}
\end{array}
\right.
\label{eq-list-A112-A111})
\end{eqnarray}
The last two values of the list (\ref{eqA112roots}) yield much bigger expressions,
useless for our purpose.

\vfill \eject 
\end{document}